\documentclass[aps,onecolumn,showpacs,superscriptaddress,groupedaddress,nofootinbib]{revtex4} 
\usepackage{graphicx}
\usepackage[sort&compress]{natbib}
\usepackage{subfigure}
\usepackage{amsmath}
\usepackage{amssymb}
\usepackage{amsfonts}
\usepackage{rotating}
\usepackage{cancel}
\usepackage{lipsum}
\usepackage{mathtools}
\usepackage{color}
\usepackage{bbm}
\usepackage{dsfont}
\usepackage{bbold}
\usepackage{multirow}
\usepackage{ulem}
\newcommand{\eps}{\epsilon}

\begin{document}

\title{Molecular picture for the $X_0(2866)$ as a $D^* \bar{K}^*$ $J^P=0^+$ state and related $1^+,2^+$ states}

\author{R. Molina}
\author{E. Oset}
\affiliation{Departamento de F\'{\i}sica Te\'orica and IFIC,
Centro Mixto Universidad de Valencia-CSIC,
Institutos de Investigaci\'on de Paterna, Aptdo. 22085, 46071 Valencia, Spain}

\begin{abstract}  
We recall the predictions made ten years ago about a bound state of $J^P=0^+$ in $I=0 $ of the $D^* \bar{K}^*$ system, which is manifestly exotic, and we associate it to the $X_0(2866)$ state reported in the recent LHCb experiment. Fine tuning the parameters to reproduce exactly the mass and width of the $X_0(2866)$ state, we report two more states stemming from the same interaction, one with $1^+$ and the other with $2^+$. For reasons of parity, the $1^+$ state cannot be observed in $D\bar{K}$ decay, and we suggest to observe it in the $D^*\bar{K}$ spectrum. On the other hand, the $2^+$ state can be observed in  $D \bar{K}$ decay but the present experiment has too small statistics in the region of its mass to make any claim. We note that measurements of the $D^*\bar{K}$ spectrum and of the $D \bar{K}$ with more statistics should bring important information concerning the nature of the $X_0(2866)$ and related ones that could be observed. 
\end{abstract}
\maketitle
\section{Introduction}
The study of mesons and baryons with more complex structure than the standard $q \bar{q}$ or $qqq$ has been a constant from the very beginning of the quark model \cite{GellMann:1964nj,Zweig:1981pd}, through works as   \cite{Jaffe:1976ig,Jaffe:1976ih,Chan:1977st,Chao:1979tg,Chao:1979mm,Hogaasen:1978jw,Strottman:1979qu,Lipkin:1987sk,Gignoux:1987cn}. Thorough recent reviews on this subject can be seen in \cite{Chen:2016qju,shilin,Karliner:2017qhf}. Some of these multiquark states are of  molecular type, meson-meson or meson-baryon, and the subject has also been thoroughly investigated. Reviews on this issue can be seen in \cite{Oller:2000ma,Oset:2016lyh,Olsen:2017bmm,Karliner:2017qhf,shilin,Guo:2017jvc}. 
  Concerning mesons, the $Z_c$ states with hidden charm and $I=1$ \cite{Ablikim:2013wzq,Ablikim:2013mio} must have four quarks, independent of whether they are molecular states of compact tetraquarks. The same can be said about the $Z_b$ states \cite{Belle:2011aa}.  Yet, the recent finding of the LHCb collaboration \cite{cerntetra} with two states of $J^P=0^+,1^-$ decaying to $D \bar{K}$ ($\bar{D}K$ in the experiment) offers us the first clear example of an exotic hadron with open heavy flavor, of type $cs \bar{u} \bar{d}$.
The states found are  
\begin{eqnarray}
&&X_0(2866) : M = 2866 \pm 7\quad \mathrm{and}\quad \Gamma = 57.2 \pm 12.9 \,\mathrm{MeV},\nonumber\\
&&X_1(2900) : M = 2904 \pm 5\quad \mathrm{ and}\quad \Gamma = 110.3 \pm 11.5 \,\mathrm{MeV}. 
\nonumber
\end{eqnarray}
    The experimental finding has already triggered theoretical work aiming at describing the states. The work of \cite{gengxie} assumes the $0^+$ state to be a bound state of $D^*\bar{K}^*$ nature. They use one boson exchange model respecting heavy quark spin symmetry (HQSS) and with reasonable parameters for the theory the $0^+$ state can be bound. They also  suggest that a $1^+$ state could also be bound, although with smaller binding, while a possible $2^+$ state is not bound. The experimental $1^-$ state does not show up in that picture, which is not surprising as vector mesons are supposed not to be molecular states \cite{sigma}. A follow up paper \cite{gxwidth} studies the decay of the $0^+$ state into $D\bar{K}$ and $D^* \bar{K}\pi$. A different method is used, following the formalism of \cite{Dong:2017gaw}, which relies upon the Weinberg compositeness condition to determine the coupling of the state to the $D^*\bar{K}^*$ component, and triangle diagrams to lead to the final decay state. 
Reasonable results for the $X_0(2866)$ width are obtained but, once again, the $X_1(2900)$ does not stand as a molecular state. 
   
    The molecular picture is also used in \cite{qianwang} using amplitudes based on HQSS and the parameters are fixed to get a bound state $X_0(2866)$. These parameters have the right size to understand also the $D_{s0}(2317)$ and $D_{s1}(2460)$ as $DK$ and $D^* K$ molecules, respectively.  In addition they get three degenerate $D^*\bar{K}^*$ states, $X_0(2866)$, one $1^+$ and one $2^+$ state with the same mass, and an extra $1^+$ state with $2722$ MeV. This degeneracy is a consequence of a strict HQSS, which we break in our approach by means of subleading HQSS terms.

    The molecular picture is again retaken in \cite{he} where the one boson exchange model is used and, with reasonable parameters, the $X_0(2866)$ can be obtained as a $D^* \bar{K}^*$ molecule, while the $X_1(2900)$ is suggested to be a bound state of $D_1 \bar{K}$. The model fitting the 
$X_0(2866)$ supports another $1^+$ bound state and a virtual $2^+$ state of $D^*\bar{K}^*$. Similar conclusions concerning the  $X_0(2866)$ state are reached in \cite{huaxing} using QCD sum rules, where the $X_0(2866)$ is favored to be a $D^* \bar{K}^*$ bound state, while the $X_1(2900)$
should be interpreted as a compact diquark-antidiquark tetraquark state. The sum rules method is also considered in \cite{zhang} where the $X_0(2866)$ could be interpreted as a diquark-antidiquark tetraquark state, within  uncertainties of $\pm200$ MeV typical of the sum rules method. On the other hand, another sum rules results of \cite{zhigang} favor the $X_0(2866)$ to be a state of axial-vector-diquark-axial-vector-antidiquark nature.

 In \cite{huang} a calculation at the quark level, using the  quark
delocalization color screening model (QDCSM) \cite{Huang:2019otd}, is used as a source of quark interaction, together with the resonating group method (RGM) \cite{kamimura} to evaluate the matrix elements,  and the molecular structure is again supported. 

     Explicit calculations for tetraquak structures based on the quark interaction have also been performed. In \cite{rosner}, a compact $cs \bar{u} \bar{d}$ tetraquark is favored for the  
$X_0(2866)$ based on a study using effective masses of quarks in heavy mesons or baryon states, and also using universal quark masses and the concept of string junctions which ascribes a mass contribution S to each QCD string junction, where the
meson has none, a baryon has one, and a tetraquark has two string junctions.  A radially excited tetraquark and orbitally excited tetraquark are also proposed as candidates for the $0^+$ and $1^-$ states, and predictions for other states are also done in \cite{weiwang} based on former work done in \cite{old}. On the other hand, an explicit tetraquark study in an extended relativized quark model done in \cite{qifang} disfavors the compact tetraquark assignment to the $X_0(2866)$. Finally, it is also interesting to quote the possibility that the observed peaks could correspond to triangle singularities as pointed out in \cite{liuxie}.  So far, the support for the 
 $D^* \bar{K}^*$ bound state nature of the $X_0(2866)$ is getting more consensus \cite{gengxie,gxwidth,Strottman:1979qu,qianwang,he,huaxing,huang} and further work and data will help to clarify the panorama in the near future. In this context it is worth exposing our point of view on the issue. 

    The first thing to point out is that in 2010, in Ref. \cite{tania}, we made neat predictions for the existence of a bound $D^* \bar{K}^*$ state with $I=0, J^P=0^+$ with mass, $2848$ MeV decaying to $D \bar{K}$ with a width of about $59$ MeV. This is in remarkable agreement with the experimental findings for the $X_0(2866)$, both in the mass and the width, without fitting any parameter to unexisting data at that time.  One may wonder how this prediction could be made, which we explain below. 

  In the first place the use of extensions of the chiral unitary approach in coupled channels \cite{Oller:2000ma} in the $D$ sector produced the $D_{s0}^*(2317)$ \cite{dani} and the $D_{s1}(2460)$ \cite{daniaxial} as molecular states of mostly $DK$ and $D^*K$ respectively. Lattice QCD calculations support this picture \cite{sasa,peardon}.  The next step would be to investigate the $D^* K^*$ states and this was done in \cite{tania}, where, among other states, the $D_{s2}^*(2573)$ state was obtained being also well described, using the fine tuning allowed for the parameters of the theory, once they are fitted to the bulk data of other states. With the input used to obtain the $D_{s2}^*(2573)$, bound states of $D^* \bar{K}^*$  nature were also obtained. The width is obtained via box diagrams with $D \bar{K}$ in the intermediate states mediated by pion exchange. In fact, an exact evaluation of the four meson loops was done. The loop was regularized with a form factor that determines the width of the states. Fine tuning of this form factor was done to get the experimental $D_{s2}^*(2573)$ width. The use of this form factor provided about $59$ MeV for the width of the $D^*\bar{K}^*$  $ 0^+$ bound state and the mass found was $2848$ MeV.

  The vector-vector interaction is much less studied than the corresponding pseudoscalar-pseudoscalar and pseudoscalar-vector ones. The reason is that for the latter ones there are available standard chiral Lagrangians in \cite{chiral,Birse:1996hd}, which are absent for the vector-vector case. Yet, the vector-vector interaction could be well taken into account by means of the local hidden gauge approach \cite{hidden1,hidden2,hidden4,nagahiro}. It is a welcome feature that the same approach also leads to the chiral Lagrangians of \cite{chiral,Birse:1996hd} assuming vector meson dominance, as found in \cite{derafael}, so it is a natural extension of the chiral Lagrangians to the vector-vector case. The unitarization in coupled channels using the local hidden gauge approach as a source of the potential leads to the chiral unitary approach for vector-vector interaction and gives rise to several dynamically generated states, molecular states, like the $f_2(1270)$ and $f_0(1370)$ in \cite{Molina:2008jw} from the $\rho \rho$ interaction and the $f_2'(1525)$, $f_0(1710)$, $K_2^*(1430)$ among others from $K^*\bar{K}^*$ and related SU(3) channels in \cite{Geng:2008gx}. Limitations of the method used in \cite{Molina:2008jw,Geng:2008gx} for very bound states were discussed in  \cite{Gulmez:2016scm,Du:2018gyn}, but the alternatives proposed were not suited for these energies as discussed in \cite{Geng:2016pmf,Molina:2019rai}. Instead, an improved method was developed in \cite{Geng:2016pmf} which actually gives very close results to those in \cite{Molina:2008jw,Geng:2008gx}. The results obtained also explain radiative decays of these resonances \cite{Nagahiro:2008um} and other decays \cite{Oset:2012zza}. 

   Confidence on one approach grows when apart from explaining known features of many states, new states are predicted that are later on found experimentally. In this sense it is worth noting that in the application of the method to study the $\rho (\omega)D^*$ interaction in \cite{Molina:2009eb} three states were found corresponding to $0^+,1^+,2^+$, the $D_0(2600)$,
$D^*(2640)$, and $D_2^*(2460)$. The $D_0(2600)$ was a prediction at that time and soon it was found in \cite{delAmoSanchez:2010vq}. 

    Since HQSS is invoked in constructing potentials for the $D^* \bar{K}^*$ system, it is worth mentioning that the local hidden gauge relies upon the exchange of vector mesons and a contact term. The dominant terms of the interaction stem from the exchange of light vectors, and this respects HQSS in an obvious way, because the heavy quarks are spectators in this exchange (see technical details in \cite{xiaoliang}). The exchange of heavy vectors and the contact terms are subleading in the HQSS counting and thus are not subject to the strict HQSS rules. Then it is interesting to note that the local hidden gauge interaction when the exchange of heavy vectors and the local terms are eliminated is the same for $J=0^+,1^+,2^+$ (see Table XI of \cite{tania}), and hence in this limit we would obtain three degenerate states, as is the case of \cite{qianwang}. When the subleading terms are kept, the degeneracy is broken and in \cite{tania}  we found three states with masses $2848$ MeV for $0^+$, $2839$ MeV for $1^+$ and $2733$ for $2^+$. 
     With the advent of the new LHCb experiment \cite{cerntetra}, we can do minor changes in the parameters of the theory to perfectly fit the mass and width of the $X_0(2866)$ and then refine the predictions for the $1^+$ and $2^+$ states. These predictions, backed by the success of the method used for the vector-vector interaction in related problems should be a stimulus to look into the experiment with improved statistics to see new peaks for the states that the theory predicts. Success in this enterprise would provide a strong backing to the molecular picture of the $X_0(2866)$ state. As to the $X_1(2900)$, with $1^-$, our approach based on the s-wave interaction of vector mesons, clearly cannot provide this state. This is the same conclusion reached in \cite{gengxie,gxwidth,qianwang,huaxing,he}.

\section{Formalism}
\begin{figure}
 \centering
 \includegraphics[scale=0.75]{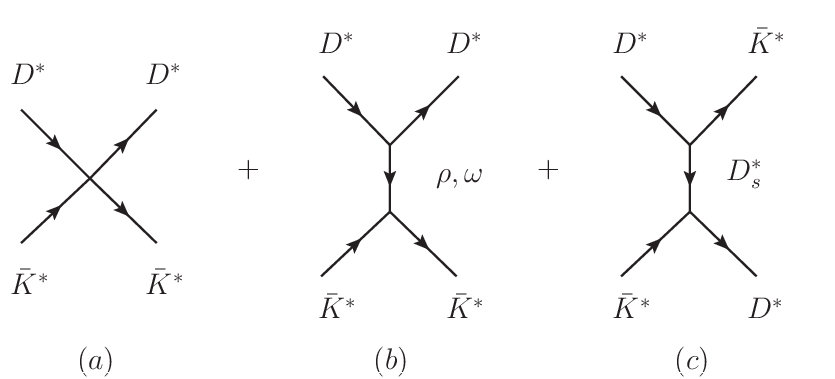}
 \caption{Feynman diagrams for the terms of the hidden local gauge approach contributing to the $D^*\bar{K}^*\to D^*\bar{K}^*$ interaction at the tree level; (a) contact term; (b) exchange of light vectors; (c) exchange of a heavy vector.}
 \label{fig:fig1}
\end{figure}

We follow the steps of \cite{tania} to describe the $V-V$ interaction for the $D^*\bar{K}^*$ case (for the conjugate state, $\bar{D}^*K^*$, the interaction would be identical). At the same time we discuss issues related to recent work on the subject. The $V-V$ interaction at tree level is given in terms of two Lagrangians of the local hidden gauge approach extended to the charm sector
\begin{eqnarray}
 &&{\cal L}_{VVVV}=\frac{1}{2}g^2\langle [V_\mu,V_\nu]V^\mu V^\nu\rangle,\nonumber\\
&& {\cal L}_{VVV}=ig\langle (V^\mu \partial_\nu V_\mu -\partial_\nu V_\mu V^\mu)
V^\nu)\rangle
\label{eq:lag}
\end{eqnarray}
where $g=M_V/2f_\pi$ ($M_V=800$ MeV, $f_\pi=93$ MeV) and $V_\mu$ is given by
\begin{equation}
\renewcommand{\tabcolsep}{1cm}
\renewcommand{\arraystretch}{2}
V_\mu=\left(
\begin{array}{cccc}
\frac{\omega+\rho^0}{\sqrt{2}} & \rho^+ & K^{*+}&\bar{D}^{*0}\\
\rho^- &\frac{\omega-\rho^0}{\sqrt{2}} & K^{*0}&D^{*-}\\
K^{*-} & \bar{K}^{*0} &\phi&D^{*-}_s\\
D^{*0}&D^{*+}&D^{*+}_s&J/\psi\\
\end{array}
\right)_\mu.
\label{eq:vmu}
\end{equation}
   The first Lagrangian in Eq. (\ref{eq:lag}) is a contact term involving four vectors. The second one is a three vector vertex which gives rise to a vector exchange diagram. The mechanisms are depicted in Fig. \ref{fig:fig1}.

The mechanism of Fig. \ref{fig:fig1} (b) corresponds to light vector meson exchange. In this mechanism the $c$ quarks are spectators and therefore, the interaction does not depend on the $c$ quark. As a consequence of that, the rules of HQSS are automatically fulfilled. Actually, the $s$ quark in $\bar{K}^*$ is also a spectator in this case and hence, one is involving only $u,d$ quarks in the process. Even if formally we can evaluate the term using the SU(4) structure of Eqs. (\ref{eq:lag}) and (\ref{eq:vmu}), only the SU(2) subgroup of it is effectively used. In fact, the vertices $D^*D^*\rho(\omega)$ can be evaluated directly using the flavor wave functions of the $\rho$ and $\omega$ without invoking any SU(4) structure \cite{sakairoca}. The diagram of Fig. \ref{fig:fig1} (c) is suppressed because of the mass of the heavy vector exchanged and is subleading in the HQSS counting, and so is the contact term of Fig. \ref{fig:fig1} (a) as shown in \cite{soler}.

The amplitudes corresponding to these three mechanisms are evaluated in terms of their polarizations and then they are projected over $J=0,1,2$ using the projector operators \cite{Molina:2008jw}
\begin{eqnarray}
&&{\cal P}^{(0)}= \frac{1}{3}\eps_\mu \eps^\mu \eps_\nu \eps^\nu\nonumber\\
&&{\cal P}^{(1)}=\frac{1}{2}(\eps_\mu\eps_\nu\eps^\mu\eps^\nu-\eps_\mu\eps_\nu\eps^\nu\eps^\mu)\nonumber\\
&&{\cal P}^{(2)}=\lbrace\frac{1}{2}(\eps_\mu\eps_\nu\eps^\mu\eps^\nu+\eps_\mu\eps_\nu\eps^\nu\eps^\mu)-\frac{1}{3}\eps_\mu\eps^\mu\epsilon_\nu\epsilon^\nu\rbrace\ .
\label{eq:projmu}
\end{eqnarray}
\begin{table}[t]
\renewcommand{\arraystretch}{1.4}
\begin{center}
\begin{tabular}{ll|rr|r}
\hline
 $J$&Amplitude&Contact & V-exchange& $\sim $ Total\\
 \hline
$0$&$D^*\bar{K}^*\to D^*\bar{K}^*$&$4g^2$&$-\frac{g^2(p_1+p_4).(p_2+p_3)}{m_{D^*_s}^2}+\frac{1}{2}g^2(\frac{1}{m_\omega^2}-\frac{3}{m_\rho^2})(p_1+p_3).(p_2+p_4)$&$-9.9g^2$\\
$1$&$D^*\bar{K}^*\to D^*\bar{K}^*$&$0$&$\frac{g^2(p_1+p_4).(p_2+p_3)}{m_{D^*_s}^2}+\frac{1}{2}g^2(\frac{1}{m_\omega^2}-\frac{3}{m_\rho^2})(p_1+p_3).(p_2+p_4)$&$-10.2g^2$\\
$2$&$D^*\bar{K}^*\to D^*\bar{K}^*$&$-2g^2$&$-\frac{g^2(p_1+p_4).(p_2+p_3)}{m_{D^*_s}^2}+\frac{1}{2}g^2(\frac{1}{m_\omega^2}-\frac{3}{m_\rho^2})(p_1+p_3).(p_2+p_4)$&$-15.9g^2$\\
\hline
\end{tabular}
\end{center}
\caption{Tree level amplitudes for $D^*\bar{K}^*$ in $I=0$. The last column shows the value of $V$ at threshold. }
\label{tab:v}
\end{table}
As a consequence, we obtain the tree level terms, $V$, shown in Table \ref{tab:v}.

In Table \ref{tab:v} we can see explicitly the contribution of the contact term, which is different for each value of $J$, and the contribution of the $D^*_s$ exchange, which is the same for $J=0,2$, but has opposite sign for $J=1$. We should note that in the absence of the subleading terms, contact and $D^*_s$ exchange, the interaction is the same for all the three cases and thus we should expect a degenerate spectrum. This is indeed what happens in \cite{qianwang}. The subleading terms are smaller than the dominant term from $\rho,\omega$ exchange, but not negligible, and they lead to a breakup of the degeneracy of the three $J$ states. 

The full amplitude is obtained by solving the Bethe-Salpeter equation with this potential,
\begin{equation}
 T=[\hat{1}-VG]^{-1}V\label{eq:bethe}
\end{equation}
with $G$ the loop function of the intermediate $D^*$ and $\bar{K}^*$. The width of the $\bar{K}^*$ is taken into account convolving the $D^*\bar{K}^*$ loop function with the $\bar{K}^*$ spectral function \cite{tania}, and $G$ is regularized in dimensional regularization by means of a subtraction constant as
\begin{eqnarray}
G_i(s)&=&{1 \over 16\pi ^2}\biggr( \alpha +\mathrm{Log}{M_1^2 \over \mu ^2}+{M_2^2-M_1^2+s\over 2s}
  \mathrm{Log}{M_2^2 \over M_1^2}\nonumber\\ 
  &+&{p\over \sqrt{s}}\Big( \mathrm{Log}{s-M_2^2+M_1^2+2p\sqrt{s} \over -s+M_2^2-M_1^2+
  2p\sqrt{s}}+\mathrm{Log}{s+M_2^2-M_1^2+2p\sqrt{s} \over -s-M_2^2+M_1^2+  2p\sqrt{s}}\Big)\biggr)\ ,
  \label{eq:gloop}
\end{eqnarray}
where $s$ is the squared c.m. energy, $p$ is the on-shell three-momentum of the two mesons, and $M_1, M_2$ the masses of the two mesons.
 Eq. (\ref{eq:bethe}) with the single channel $D^*\bar{K}^*$, and using the value of $\alpha$ necessary to reproduce the $D_{s2}(2573)$, gives rise to a bound $D^*\bar{K}^*$ state with $0^+$ around $2848$ MeV.

\begin{figure}
 \centering
 \includegraphics[scale=0.75]{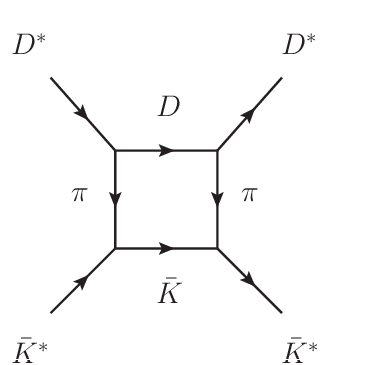}
 \caption{Box diagram accounting for the width of the $\bar{D}^*\bar{K}^*$ state decaying to $D\bar{K}$.}
 \label{fig:fig2}
\end{figure}

In the absence of the $K^*$ width the state has zero width. The convolution with the $\bar{K}^*$ spectral function gives rise to a tiny width, which is not the one observed. The experiment sees the state in the $\bar{D}K$ ($D\bar{K}$ in our case) decay. This decay was also studied in \cite{tania} by means of the box diagram that leads to this decay, as shown in 
Fig. \ref{fig:fig2}.

The amplitude given by the diagram of Fig. \ref{fig:fig2} is added to the potential discussed above and iterated with the Bethe Salpeter equation to obtain bound states which now decay into $D\bar{K}$. It is interesting to see that, since $D$ and $\bar{K}$ have spin zero, we need $L=0,1,2$ for these intermediate states to match the $J=0,1,2$ of the states obtained. Since all $V-V$ states obtained have positive parity in the $s$-wave that we study, only $J=0,2$ decay into $D\bar{K}$, and the state with $J=1$ cannot. This means that the $J=1$ state that we predict cannot be seen in the experiment of \cite{cerntetra}, but $J=0,2$ could be. The $J=1$ state can be seen in $D^*\bar{K}$ decay, which requires an anomalous vertex and should be suppressed. Indeed, this is what happens for the $D^*(2640)$ state that we obtained with $J=1$ in \cite{Molina:2009eb}, with a small width compared to the $D_0(2600)$ and $D^*_2(2460)$, which have a large width. In the PDG \cite{pdg}, the widths of $D_0(2600)$, $D^*(2640)$, $D^*_2(2460)$ are respectively $139$ MeV, $<15$ MeV, $46.7$ MeV. 

The evaluation of the diagram in Fig. \ref{fig:fig2} requires a regularizing form factor. In \cite{tania} we took a form factor from \cite{navarra}
\begin{eqnarray}
F(q)=e^{((p_1^0-q^{0})^2-\vec{q}\,^2)/\Lambda^2}\ ,
\label{eq:formfactorprim}
\end{eqnarray}
with $q_0=(s+m^2_D-m^2_K)/2\sqrt{s}$ and $\Lambda$ of the order of $1-1.2$ GeV, and $p_1^0$ is the initial $D^*$ energy, $p_1^0\simeq m_{D^*}$\footnote{We take advantage to note a typo in Eq. (30) of Ref. \cite{tania} where $q_0^2$ in the exponent of $F(q)$ should be $(k^0_1-q^0)^2$, which is the actual form factor used in \cite{tania} and here in Eq. (\ref{eq:formfactorprim}).}. The amplitude of Fig. \ref{fig:fig2} gave negligible contribution to the real part of the energy of the bound states, but provides the width for $D\bar{K}$ decay. Choosing $\alpha=-1.6$ (with $\mu=1500$ MeV) and $\Lambda=1200$ MeV, we obtained in \cite{tania} the results of Tables \ref{tab:tab2} and \ref{tab:tab3}.
\begin{table}[t]
\renewcommand{\arraystretch}{1.2}
 \begin{center}
  \begin{tabular}{lrrrr}
  \hline
   $I(J^P)$&$M[\mathrm{MeV}]$ & $\Gamma[\mathrm{MeV}]$ & Channels & state\\
   \hline
   $0(2^+)$&$2572$&$23$&$D^*K^*,D_s^*\phi,D^*_s\omega$&$D_{s2}(2572)$\\
   $0(1^+)$&$2707$&-&$D^*K^*,D^*_s\phi,D^*_s\omega$&?\\
   $0(0^+)$&$2683$&$71$&$D^*K^*,D^*_s\phi,D^*_s\omega$&?\\
   \hline
  \end{tabular}
\caption{States with $C=1,S=1,I=0$ from \cite{tania}. }
\label{tab:tab2}
 \end{center}
\end{table}

\begin{table}[t]
\renewcommand{\arraystretch}{1.2}
 \begin{center}
  \begin{tabular}{lrrrr}
  \hline
   $I(J^P)$&$M[\mathrm{MeV}]$ & $\Gamma[\mathrm{MeV}]$ & Channels & state\\
   \hline
   $0(2^+)$&$2733$&$36$&$D^*\bar{K}^*$&?\\
   $0(1^+)$&$2839$&-&$D^*\bar{K}^*$&?\\
   $0(0^+)$&$2848$&$59$&$D^*\bar{K}^*$&$X_0(2866)$\\
   \hline
  \end{tabular}
\caption{States with $C=1,S=-1,I=0$ from \cite{tania}.}
\label{tab:tab3}
 \end{center}
\end{table}
The masses and widths are evaluated normally from the poles in the second Riemann sheet. When the convolution in the $G$-function is done it is common to obtain them from the $|T|^2$ plot instead.

\begin{figure}
 \centering
 \includegraphics[scale=1]{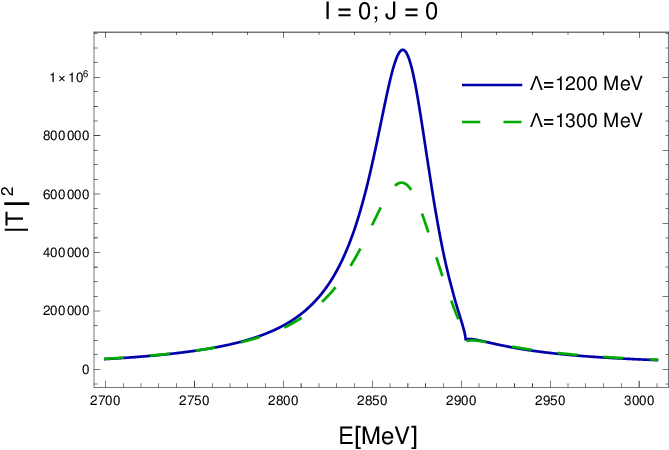}
 \caption{$|T|^2$ for $J=0$ and $C=1,S=-1,I=0$.}
 \label{fig:fig3}
\end{figure}

\begin{figure}
 \centering
 \includegraphics[scale=1]{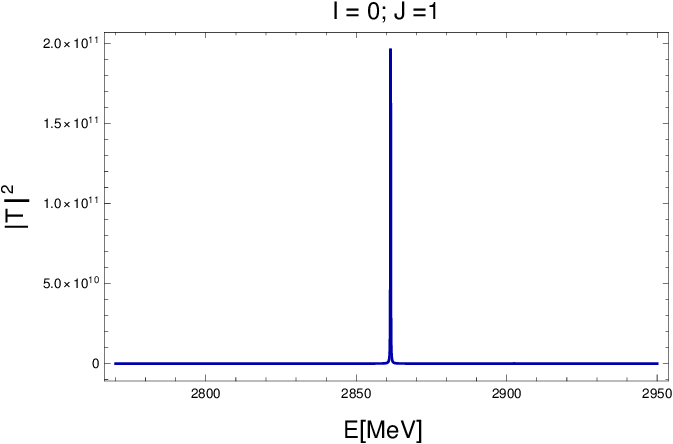}
 \caption{$|T|^2$ for $J=1$ and $C=1,S=-1,I=0$ (the small width comes from the convolution of the $\bar{K}^*$ width).}
 \label{fig:fig4}
\end{figure}

\begin{figure}
 \centering
 \includegraphics[scale=1]{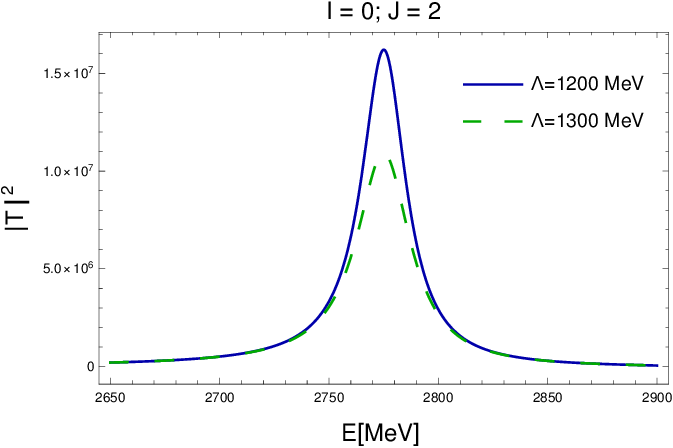}
 \caption{$|T|^2$ for $J=2$ and $C=1,S=-1,I=0$.}
 \label{fig:fig5}
\end{figure}

In view of the new experimental data of \cite{cerntetra} we do a fine tuning of the $\alpha$ and $\Lambda$ parameters to obtain the experimental numbers. In Figs. \ref{fig:fig3}, \ref{fig:fig4} and \ref{fig:fig5}, we show $|T|^2$ for the $D^*\bar{K}^*$ states with $L=0^+,1^+,2^+$.
By inspecting these figures we obtain $M(0^+)=2866$ MeV, $\Gamma=57$ MeV with $\alpha=-1.474$, $\Lambda=1300$. The new results for the three $D^*\bar{K}^*$ states are shown in Table \ref{tab:tab4}.
\begin{table}[t]
\renewcommand{\arraystretch}{1.2}
\begin{center}
 \begin{tabular}{lrrrr}
 \hline
  $I(J^P)$&$M[\mathrm{MeV}]$&$\Gamma[\mathrm{MeV}]$& Coupled channels & state\\
  \hline
  $0(2^+)$& $2775$ & $31$ & $D^*\bar{K}^*$ &?\\
  $0(1^+)$& $2861$ & $-$ & $D^*\bar{K}^*$ &?\\
  $0(0^+)$&$2866$ & $57$& $D^*\bar{K}^*$ &$X_0(2866)$\\
  \hline
 \end{tabular}
\end{center}
\caption{$D^*\bar{K}^*$ states obtained by fine tuning of the free parameters.}
\label{tab:tab4}
\end{table}
As we can see, the parameters are very close to those used in \cite{tania} since only a small difference in the mass and width has to be accommodated. As a consequence, the results for the $1^+,2^+$ states, which are predictions of our theoretical framework, are very similar to those obtained in \cite{tania} (see Table \ref{tab:tab3}).

As discussed above, the $J^P=1^+$ state cannot be seen in the $D\bar{K}$ spectrum (or $D^-K^+$ in the experiment \cite{cerntetra}) but the $2^+$ state could. If one looks into the $D^-K^+$ spectrum of \cite{cerntetra}, there is not enough statistics in that region to make any claim so far. By simple analogy to the $D_{s2}(2572)$ state from $D^*K^*$, the $2^+$ state from $D^*\bar{K}^*$ should also exist, and this simple observation, in addition to our predictions, should give incentives to look for this additional state with more statistics. The search for the $1^+$ state in the $D^*\bar{K}$ spectrum would be a complement to this search to find the whole family of exotic $D^* \bar{K}^*$ molecular states. In the next subsection we evaluate the width of this state and include also the new width for the $2^+$ state. Note that with $J^P=0^+$, the $D^*\bar{K}$ intermediate requires $L=1$ to match the angular momentum, but this gives negative parity and hence the $0^+$ state cannot decay into this channel.

\subsection{The width of the $I=0;J=1$ state}
\begin{figure}
\centering
 \includegraphics[scale=0.6]{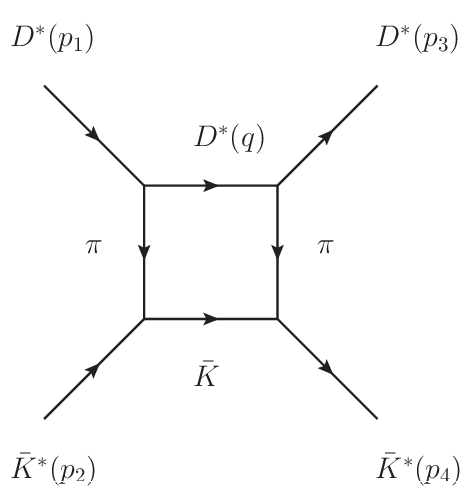}
 \caption{Box diagram containing the $D^*\bar{K}$ decay channel. }
 \label{fig:dskbox}
\end{figure}

\begin{figure}
\centering
 \includegraphics[scale=1]{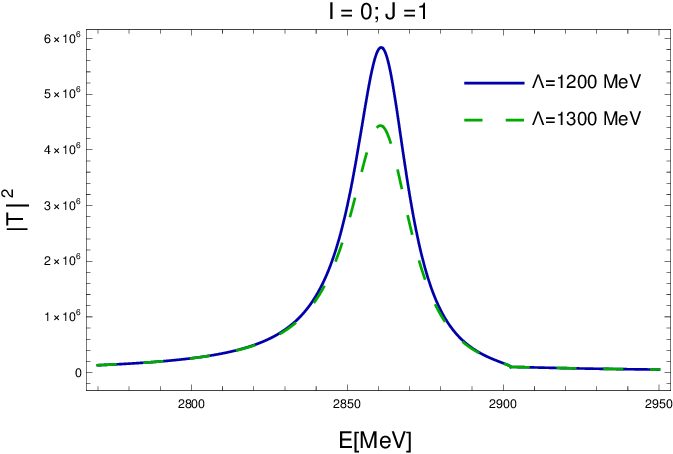}
 \caption{$\vert T\vert^2$ for $J=1,C=1$ and $S=-1,I=0$.}
 \label{fig:statej1decay}
\end{figure}

As discussed above, the $J=1$ state does not decay into $D\bar{K}$. It can decay into $D^*\bar{K}$ and $D\bar{K}^*$, but given the much larger phase space for $D^*\bar{K}$ this is the preferred channel, and we evaluate this width here corresponding to the diagram of Fig. \ref{fig:dskbox}.
One needs two Lagrangians to evaluate it, the anomalous term \cite{bramon,luis},
\begin{equation}
 {\cal L}=\frac{iG'}{\sqrt{2}}\eps^{\mu\nu\alpha\beta}\langle \delta_\mu V_\nu \delta_\alpha V_\beta P\rangle
\end{equation}
with $G'=\frac{3g'}{4\pi^2f};g'=-\frac{G_Vm_\rho}{\sqrt{2}f^2}$, $G_V\simeq 55$ MeV, $V$ the vector field matrix of Eq. (\ref{eq:vmu}), and $P$ the analogous matrix for the pseudoscalar fields (see Eq. (12) of Ref. \cite{dani}). Note that since in Fig. \ref{fig:dskbox} one is exchanging a $\pi$ meson, the $c$ quarks of $D^*$ act as spectators and the use of SU(4) is essentially formal, only its SU(3) subgroup contents is effectively used \cite{sakairoca}. 
  The other Lagrangian is the one for $V\to PP$ decay ($\bar{K}^*\to  K\pi$ here),
  \begin{equation}
   {\cal L}_{VPP}=-ig\langle [ P,\partial_\mu P]V^\mu\rangle 
  \end{equation}
  with $g=m_V/2f$ ($M_V=800$ MeV, $f=93$ MeV), with the phase convention for doublets $(D^{*+},-D^{*0})$, $(\bar{K}^{*0},-K^{*-})$, $(\bar{K}^0,-K^-)$, the $I=0$ states are written,
  \begin{eqnarray}
  && \lvert I=0,D^*\bar{K}^*\rangle=\frac{1}{\sqrt{2}}(D^{*+}\bar{K}^{*0}-D^{*+}K^{*-})\nonumber\\
  && \lvert I=0, D^*\bar{K}\rangle=\frac{1}{\sqrt{2}}(D^{*0}\bar{K}^0-D^{*+}K^-)
  \end{eqnarray}
  One can then evaluate the amplitude for the diagram of Fig. \ref{fig:dskbox} assuming zero three momenta for the external vectors with the result,
  \begin{eqnarray}
  && -it=\frac{9}{2}(G'gm_{D^*})^2\int\frac{d^4q}{(2\pi)^4}\eps^{ijk}\eps^{i'j'k'}\left( \frac{1}{(p_1-q)^2-m_\pi^2+i\eps}\right)^2\frac{1}{q^2-m^2_{D^*}+i\eps}\frac{1}{(p_1+p_2-q)^2-m^2_K+i\eps}\nonumber\\
  &&\times \eps^{j(1)}\eps^{m(2)}\eps^{k(3')}q^iq^m\eps^{j'(1)}\eps^{m'(4)}\eps^{k'(3')}q^{i'}q^{m'}F^4(q)
  \label{eq:tdskp}
  \end{eqnarray}
with $3'$ standing for the intermediate $D^*$ state and $F(q)$ the form factor of Eq. (\ref{eq:formfactorprim}). The operator $\partial_\alpha $ acts on the external $D^*$ and gives $-im_{D^*}\delta_{\alpha 0}$, and the other inidices in $\eps^{\mu\nu\alpha\beta}$ are then spatial, leading to the structure of Eq. (\ref{eq:tdskp}). One also has for the sum over internal polarizations, $\eps^{k(3')}\eps^{k'(3')}=\delta_{kk'}$. The $d^4q$ integration is done with $dq^0$ performed analytically and $d^3q$ integration numerically. This is done explicitly in Ref. \cite{Molina:2009eb} but, given the fact that the real part of the box is found negligible compared to the local hidden gauge dominant terms, we concentrate here only in the imaginary part. This is simplified by noticing that $D^*\to D^*\pi$ cannot proceed physically and the only cut leading to an imaginary part in Fig. \ref{fig:dskbox} is the $D^*\bar{K}$ cut, where the $D^*$ and $\bar{K}$ are placed on-shell. This leads to
\begin{eqnarray}
 &&t=\frac{9}{2}(G'gm_{D^*})^2\int\frac{d^3q}{(2\pi)^3}(\delta_{ii'}\delta_{jj'}-\delta_{ij'}\delta_{i'j})\left(\frac{1}{(p_1-q)^2-m_\pi^2}\right)^2\frac{1}{2\omega^*(q)}\frac{1}{2\omega(q)}\frac{1}{\sqrt{s}-\omega^*(q)-\omega(q)+i\eps}\nonumber\\
 &&\times \eps^{j(1)}\eps^{m(2)}\eps^{j'(3)}\eps^{m'(4)}q^iq^mq^{i'}q^{m'}F^4(q)
 \label{eq:tdsk}
\end{eqnarray}
with $\omega^*(q)=\sqrt{m^2_{D^*}+\vec{q}\,^2}$, $\omega(q)=\sqrt{m^2_K+\vec{q}\,^2}$, $p_1^0=m_{D^*}$, $q^0=\omega^*(q)$. Taking now into account that,
\begin{eqnarray}
 \int \frac{d^3q}{(2\pi)^3}f(\vec{q}\,^2)q^iq^mq^{i'}q^{m'}=\frac{1}{15}\int \frac{d^3q}{(2\pi)^3}f(\vec{q}\,^2)\vec{q}\,^4(\delta_{im}\delta_{i'm'}+\delta_{ii'}\delta_{mm'}+\delta_{im'}\delta_{m'i})\ ,
\end{eqnarray}
one gets a particular combination of the four $\eps$ polarizations,
\begin{eqnarray}
 4\eps^{j(1)}\eps^{m(2)}\eps^{j(3)}\eps^{m(4)}-\eps^{j(1)}\eps^{j(2)}\eps^{m(3)}\eps^{m(4)}-\eps^{j(1)}\eps^{m(2)}\eps^{m(3)}\eps^{j(4)}\ ,
\end{eqnarray}
which using the spin projectors ${\cal P}^{(0)}$, ${\cal P}^{(1)}$, and ${\cal P}^{(2)}$ of Eq. (\ref{eq:projmu}) can be cast into
\begin{equation}
 5 {\cal P}^{(1)}+3{\cal P}^{(2)}\ ,
\end{equation}
and zero component of ${\cal P}^{(0)}$, in agreement with our former argument that $J=0$ requires $L=1$ in the intermediate $D^*\bar{K}$ state and violates parity. The imaginary part of Eq. (\ref{eq:tdsk}) is readily obtained and we finally obtain for $J=1$,
\begin{eqnarray}
\mathrm{Im}t=-\frac{3}{2}\frac{1}{8\pi}(G'gm_{D^*})^2q^5\left(\frac{1}{(m_D^*-\omega^*(q))^2-\omega^2(q)}\right)^2\frac{1}{\sqrt{s}}F^4(q)\ ,
\end{eqnarray}
where we have used $q^0=\omega^*(q)$, and $q=\frac{\lambda^{1/2}(s,m^2_{D^*},m^2_{K})}{2\sqrt{s}}$. For $J=2$ we have the same formula replacing $3/2$ by $9/10$. The potential $\delta V= i\mathrm{Im}t$ is added to our former potential $V_{D^*\bar{K}^*,D^*\bar{K}^*}$ of the local hidden gauge approach and the Bethe Salpeter equation is solved again. The $\vert T\vert^2$ magnitude for the $J=1$ state is now plotted in Fig. \ref{fig:statej1decay}, from where we see that the width of the state is about $20$ MeV. The extra width obtained from $D^*\bar{K}$ decay for the $J=2$ state is about $7$ MeV. With this information we complete Table \ref{tab:tab4} in Table \ref{tab:tab5}.

\begin{table}[t]
\renewcommand{\arraystretch}{1.2}
\begin{center}
 \begin{tabular}{lrrrr}
 \hline
  $I(J^P)$&$M[\mathrm{MeV}]$&$\Gamma[\mathrm{MeV}]$& Coupled channels & state\\
  \hline
  $0(2^+)$& $2775$ & $38$ & $D^*\bar{K}^*$ &?\\
  $0(1^+)$& $2861$ & $20$ & $D^*\bar{K}^*$ &?\\
  $0(0^+)$&$2866$ & $57$& $D^*\bar{K}^*$ &$X_0(2866)$\\
  \hline
 \end{tabular}
\end{center}
\caption{As in Table \ref{tab:tab4} but including the width of the $D^*K$ channel.}
\label{tab:tab5}
\end{table}

\section{Conclusions}
  The observation in \cite{cerntetra} of two mesons in the $D^-K^+$ spectrum, with $cs \bar{u} \bar{d}$ content (for $D^+K^-$), provides the first example of an open heavy flavor exotic state.  The discovery has immediately triggered a response from the theoretical community, with different suggestions for the interpretation of these states. One of the most promising interpretations is that the $X_0(2866)$ is a molecular state of $\bar{D}^* K^*$ nature, while the $X_1(2900)$ is not easy to accommodate in that picture. In this paper we have recalled that a prediction of a bound  $D^*\bar{K}^*$ state with $0^+$ was already been done in \cite{tania}, being in remarkable agreement with the mass and width of the $X_0(2866)$. We have taken the opportunity to discuss that idea to the light of developments done after the theoretical paper \cite{tania} in connection with the data of \cite{cerntetra} and recent theoretical papers done after the experimental discovery. One of the issues is heavy quark spin symmetry, that has become fashionable in the study of heavy quark systems. We have shown that the approach followed in \cite{tania} respects HQSS. Indeed, the formalism followed to deal with the vector-vector interaction is the local hidden gauge approach, where one has a contact term and other terms stemming from the exchange of vector mesons. The dominant terms come from the exchange of light vector mesons, and since in this case the heavy quarks are spectators in the process, the amplitudes do not depend upon them and HQSS is automatically fulfilled. Yet, even if subdominant in the heavy quark counting, the contact term and the exchange of heavy vectors are not negligible and have as a consequence the splitting of the $J=0,1,2$ states generated in $s$-wave, which are degenerate in the strict heavy quark limit. 
   We have taken advantage of the measurements in \cite{cerntetra} to fine tune the two parameters of the model such as to adjust exactly the mass and the width of the $0^+$ state to the $X_0(2866)$, and then made predictions for the $1^+$ and $2^+$ states.  The $1^+$ state cannot be seen in the $D \bar{K}$ spectrum because of parity reasons. We suggest to look at it in the $D^*\bar{K}$ spectrum and we have evaluated its width.
   As to the $2^+$ state, it can be seen in the $D \bar{K}$ spectrum, but it falls in a  region of energies where the experiment has small statistics. Our results and the existence of the closely related $D_{s2}^*(2573)$ state, which stems from the $D^* K^*$ interaction, should provide an incentive to look for these predicted states. Their observation would give a strong support to the molecular picture for the $X_0(2866)$ state and related ones that could come from the experiment. 
\section{Acknowledgments}
R. M. acknowledges support from the CIDEGENT program with Ref. CIDEGENT/2019/015 and from the spanish national grant PID2019-106080GB-C21.
This work is also partly supported by the Spanish Ministerio de Economia y Competitividad and European FEDER funds under Contracts No. FIS2017-84038-C2-1-P B and No.  FIS2017-84038-C2-2-P B. This project has received funding from the European Union’s Horizon 2020 research and innovation programme under grant agreement No. 824093 for the STRONG-2020 project.

\bibliography{biblio}

\end{document}